\documentclass[apj]{emulateapj}
\usepackage[colorlinks,linkcolor={blue},citecolor={blue},urlcolor={red}]{hyperref}
\usepackage{epsfig,graphicx,natbib,amsmath,amsfonts,amssymb}
\usepackage{color}

\newcommand{\dindex}{D$_n$(4000)}
\newcommand{\nuvr}{NUV$-r$}
\newcommand{\hd}{H$\delta$}
\newcommand{\hda}{\hd$_A$}
\newcommand{\ewhda}{EW(\hda)}
\newcommand{\ha}{H$\alpha$}
\newcommand{\hae}{\ha}
\newcommand{\ewhae}{EW(\hae)}
\newcommand{\lgewhae}{$\log_{10}$\ewhae}

\newcommand{\sersic}{S\'{e}rsic}

\newcommand{\HI}{H{\sc{I}}}

\newcommand{\myemail}{\email{ecwang16@ustc.edu.cn(EW), xkong@ustc.edu.cn(XK)}}

\shorttitle{The Peculiar HI Morphology of NGC 6145}
\shortauthors{Wang et al.}

\graphicspath{{fig/}}

\begin{document}

\title{The Peculiar Filamentary HI Structure of NGC 6145}
\author{
Enci Wang\altaffilmark{1,2},
Jing Wang\altaffilmark{3},
Xu Kong\altaffilmark{1},
Fulai Guo\altaffilmark{2},
Lin Lin\altaffilmark{2},
Guobin Mou\altaffilmark{1},
Cheng Li\altaffilmark{4,2},
Ting Xiao\altaffilmark{2}
} \myemail

\altaffiltext{1}{CAS Key Laboratory for Research in Galaxies and Cosmology, Department of Astronomy, 
University of Science and Technology of China, Hefei, Anhui 230026, China}
\altaffiltext{2}{Key Laboratory for Research in Galaxies and Cosmology, Shanghai Astronomical Observatory, Chinese Astronomical Society, 80 Nandan Road,
Shanghai 200030, China}
\altaffiltext{3}{Australia Telescope National Facility, CSIRO Astronomy and Space Science, PO box 76, Epping, NSW 1710, Australia}
\altaffiltext{4}{Tsinghua Center of Astrophysics \& Department of Physics, Tsinghua University, Beijing 100084, China}

\begin{abstract}   %
 In this paper, we report the peculiar \HI\ morphology of the cluster spiral galaxy NGC 6145,
 which has a 150 kpc \HI\ filament on one side that is nearly parallel to its major axis. 
This filament is made up of several \HI\ clouds and the diffuse \HI\ gas between them, with no optical counterparts. 
 We compare its \HI\ distribution with other one-sided \HI\ distributions in the literature, 
 and find that  the overall \HI\ distribution is very different from the typical tidal and ram-pressure
  stripped \HI\ shape, and its morphology is inconsistent with being a pure accretion event.
   Only $\sim$30\% of the total \HI\ gas is anchored on the stellar disk, 
  while most of \HI\ gas forms the filament in the west. 
  At a projected distance of 122 kpc, we find a massive elliptical companion (NGC 6146) with extended 
  radio emission, whose axis points to an \HI\ gap in NGC 6145.
 The velocity of the \HI\ filament shows an overall light-of-sight motion 
 of 80 to 180 km s$^{-1}$ with respect to NGC 6145. 
Using the long-slit spectra of NGC 6145 along its major stellar axis,
 we find that some outer regions show enhanced 
star formation, while in contrast, almost no star formation activities are found in its center ($<$2 kpc). 
Pure accretion, tidal or ram-pressure stripping is not likely to 
produce the observed \HI\ filament. An alternative explanation is the jet-stripping from NGC 6146, 
although direct evidence for a jet-cold gas interaction has not been found. 

\end{abstract}


\section{Introduction}
\label{sec:introduction}
  
 Observationally, the \HI\ distribution of galaxies measured at 21 cm is often irregular
  and more extended than their optical counterparts.
 According to the \HI\ mappings of nearby galaxies, such as Westerbork observations of neutral Hydrogen
in Irregular and SPiral galaxies \citep[WHISP;][]{Swaters-02},  the \HI\ morphology of galaxies 
exhibits a wide variety of features, such as \HI\ tails, bridges and extra-planar \HI\ gas \citep{Fraternali-02,
 Oosterloo-Fraternali-Sancisi-07}.
  In galaxy clusters, a series of physical processes can influence or determine \HI\ morphologies, such as
   tidal stripping \citep{Farouki-Shapiro-81, Mayer-06, Haynes-11}, ram-pressure stripping \citep[e.g.][]{Moore-96, Mayer-06}, and interaction with close companions \citep{Hibbard-Yun-99, Okamoto-15}.
  However, the dominant physical process shaping the \HI\ morphology varies from case to case.
  
  Tidal stripping, which commonly occurs in galaxy groups and clusters, is quite efficient at removing gas, 
  dust and stars from galactic halos \citep{Farouki-Shapiro-81, Icke-85, Mayer-06, Haynes-11}, 
 and especially loose peripheral or extra-planar \HI\ gas \citep{Valluri-Jog-90}, 
 often associated with enhanced 
  star formation activities in galactic centers \citep{Hummel-90, Li-08, Smith-Davies-Nelson-10}. 
 As shown in many archetypal examples from observations, typical tidally stripped \HI\ morphology 
exhibits multiple tidal tails or bridges \citep{Hibbard-Vacca-Yun-00, Duc-00, Okamoto-15}, 
 such as the 180 kpc \HI\ tidal tail in merger ARP 299
\citep{Hibbard-Yun-99}, and \HI\ bridges in minor merger system of M81, M82 and NGC 3077 \citep{Yun-Ho-Lo-94},
usually as well as disturbed optical morphologies \citep{Schombert-88, Kormendy-Djorgovski-89}. 
\cite{Haynes-11} examined 199 ALFALFA objects that have
 no optical counterparts in SDSS, and found 75\% of them are located in fields where galaxies
 of similar redshifts are found, suggesting that they are likely to be related to tidal-debris fields.

  In contrast to tidal stripping, ram-pressure stripping only acts on the interstellar medium (ISM) 
  when satellite galaxies are falling into clusters. 
  A typical ram-pressure stripped \HI\ morphology shows a head-tail shape 
  \citep{Gavazzi-95,vanGorkom-03, Kenney-vanGorkom-Vollmer-04, Lee-Chung-15}. 
 By mapping the \HI\ gas distribution of galaxies in Virgo cluster, \cite{Chung-09} found 
  a remarkable number of galaxies with long, one-sided \HI\ tails pointing away from M87, which were 
  proposed to be the effect of ram-pressure stripping, in the sense that
  these galaxies are falling into the Virgo core for the first time. 
 By modeling pressure variations in the ISM during the ram-pressure stripping process, 
 \cite{Fujita-98} and \cite{Fujita-Nagashima-99} found that star formation activities can increase by 
 up to a factor of two on short timescales ($\sim$ 10$^8$ yr) in rich clusters, while the removal of 
 the \HI\ gas reservoir leads to a suppression of star formation on longer timescales.
  
 In this paper, we report a galaxy (NGC 6145) from the ``Bluedisk'' project \citep{Wang-13, Wang-15}, 
 which hosts a one-sided \HI\ filament of 150 kpc length in projected image. 
 The Bluedisk project has mapped the \HI\ emission of a sample of 23 galaxies with unusually 
 high \HI\ mass fractions, as well as a similar-sized sample of control galaxies \citep{Wang-13}. 
 NGC 6145 is from the control sample, which has a normal \HI\ mass fraction, and 
  it resides in the Abell cluster A2197 \citep{Wrobel-88},
  whose halo mass is 10$^{14.6}M_{\odot}$ \citep{Yang-07}; i.e., about one third of Virgo
  cluster \citep{Fouque-01}.   
 Three companions at similar redshifts are found within the projected distance of 150 kpc from NGC 6145. 
The \HI\ shape of NGC 6145 does not exhibit typical \HI\ tidal tails or head-tail features, indicating that
 it is not typical tidal or ram-pressure stripped galaxy. No optical counterparts 
are found to the \HI\ filament, which is different to typical interactions or mergers.  
This excited our interest to explore the physical origin of the 150 kpc \HI\ filament. 
Thus, in this work we analyzed the \HI\ properties, local environment, and star-formation status 
of NGC 6145, which provide clues to its origin.  

The remainder of this paper is organized as follows. 
 In Section 2, we briefly introduce the data we used, including the 
 \HI\ datacube and the optical long-slit spectra along the major axis of NGC 6145. 
 In Section 3, we compare the \HI\ morphology of NGC 6145 with some previous findings of
 one-sided \HI\ distribution of galaxies. We also investigate the local environment of NGC 6145,  
 the kinematics of its \HI\ gas, and the star formation history along its major axis. 
In Section 4, we summarize our results and discuss several possible mechanisms that may
 account for the observed \HI\ filament.   
Throughout this paper, all the distance-dependent parameters were computed 
assuming a flat $\Lambda$CDM cosmological model with $\Omega_m=0.27$,
 $\Omega_\Lambda=0.73$ and $h=0.7$.
The redshift of NGC 6145 is 0.0287, which corresponds to a luminosity 
  distance of 126 Mpc from us. 

\section{Data}
\label{sec:data}

NGC 6145 is one of the 50 targeted galaxies in ``Bluedisk'' project. 
Both \HI\ interferometry and optical long-slit spectral data were obtained.
  A full description of observations and data reductions of these datasets were presented
by \cite{Wang-13} and \cite{Carton-15}. 
Therefore, we give only a brief introduction of these procedures in this section.

\subsection{\HI\ Observations and data reduction}

  The 21 cm emission of NGC 6145 was observed with Westerbork Synthesis Radio Telescope (WSRT) 
  in March 2012 with an on-source integration time of 12 hours. 
  The raw data cube was produced using the pipeline described in \cite{Serra-Jurek-Floer-12}, 
  which is based on the MIRIAD reduction package 
  \citep{Sault-Teuben-Wright-95}. In this paper, the \HI\ cube was built with a robust weighting of 0.4,
   which provided a suitable compromise between sensitivity and resolution. 

The velocity resolution is $24.8$ km s$^{-1}$ (FWHM) and
the typical beam has a half-power beamwidth (HPBW) of $16.1 \times 25.3$ arcsec$^2$.
The \HI\ cube covers a redshift range of $\Delta z=0.0063$, corresponding to a velocity range of  
$\sim$1900 km s$^{-1}$, and has a size of 1$^\circ$ on each side, corresponding to a physical scale
 of $\sim$2.0 Mpc at a redshift $z=0.0287$ (the redshift of NGC 6145).
This means that close companions of NGC 6145 were also observed by WSRT at the same time.

We generated a two-dimensional \HI\ total-intensity map of NGC 6145 in two steps. First, we identified 3-dimensional
 regions with 21 cm emission with a smoothing and clipping algorithm. We then added all the detected \HI\ emission 
along the velocity direction. We also estimated errors for all non-zero pixels in the \HI\ intensity map.

\subsection{Optical long-slit spectroscopy}

\subsubsection{Observation and data reduction}
The optical long-slit spectrum of NGC 6145 was obtained using the 
Intermediate dispersion Spectrograph and Imaging System (ISIS) on the 
4.2 m William Herschel Telescope (WHT). Details of data and their full reduction were
described by \cite{Carton-15}, although we have briefly summarized the relevant details here. 

The spectrograph was set up to observe over the wavelength ranges 
3700 - 5300 \AA\ and 5750 - 7200 \AA\ simultaneously, with a spectral resolution 
that is effectively described by a Gaussian with a $\sim$1.7 \AA\ FWHM. The 
observation was performed on the night of 2nd May 2013,  and the seeing 
conditions varied between 0.7 and 1.4 arcsec. Note that this is 
significantly smaller than the 3 arcsec slit width. The total 
integration time was 1h (3$\times$1200s).

The spectroscopic slit was oriented to pass through the center
and along the major axis of the galaxy (left panel of Figure 1). 
The position angle (PA) is 4$^{\circ}$, as determined from the SDSS image.

\subsubsection{Spectral fitting}

\begin{figure*}
  \begin{center}
    \epsfig{figure=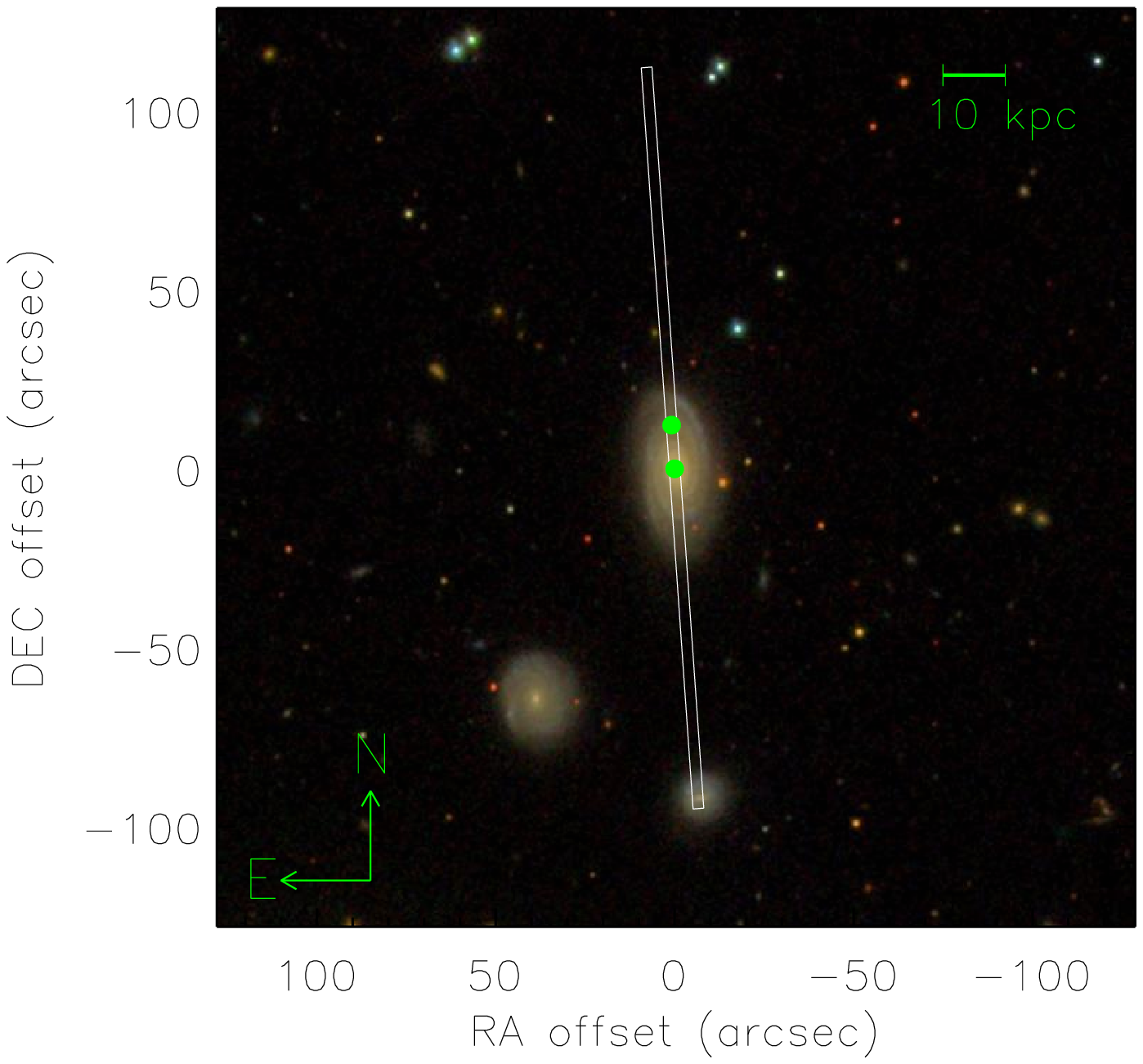,clip=true,width=0.33\textwidth}
    \epsfig{figure=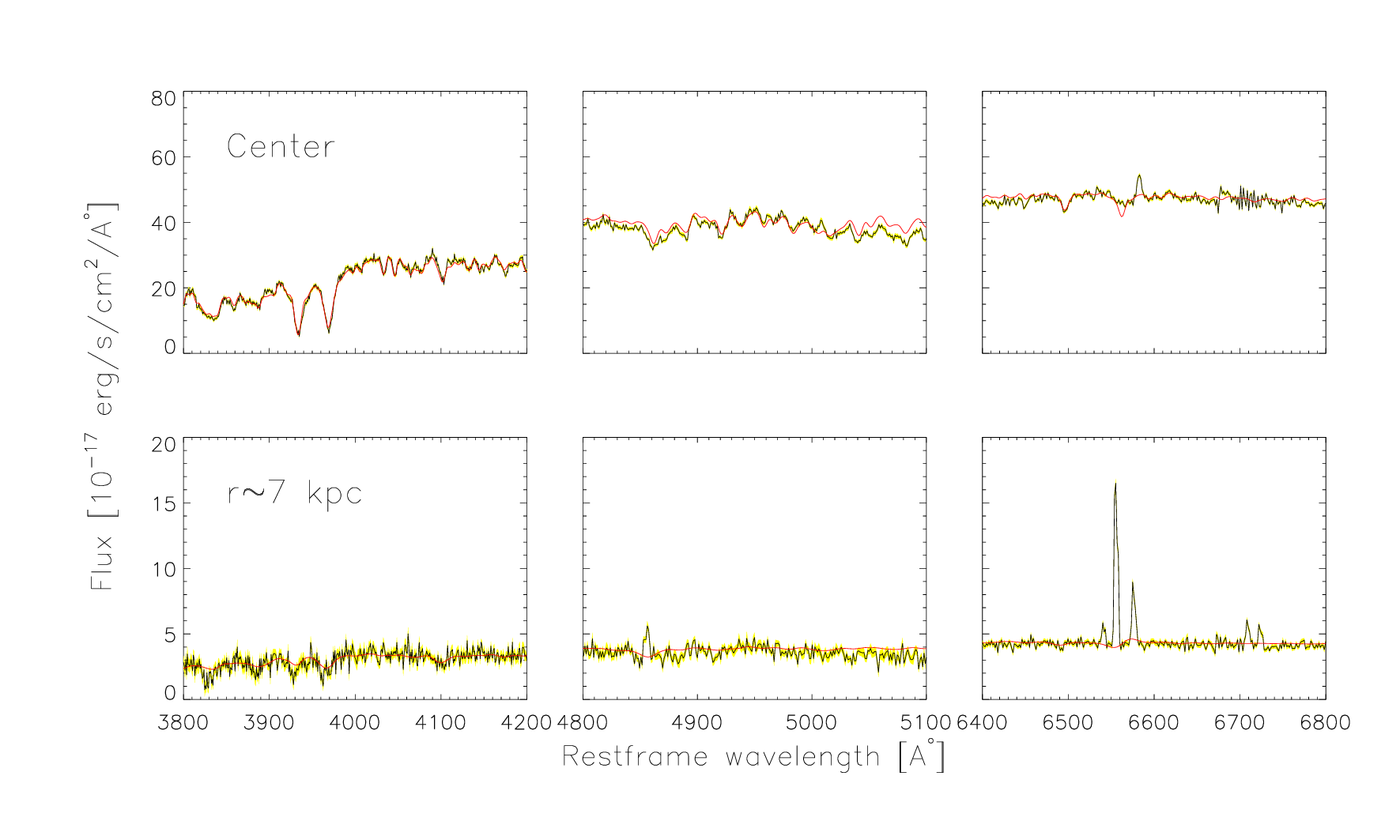,clip=true,width=0.66\textwidth}
  \end{center}
  \caption{ Left panel: WHT ISIS slit position (shown as the narrow, white box) overlaid on the SDSS optical image of NGC 6145.
The orientation of the slit is along the major axis of the galaxy. Right panels: Two examples of spectral fitting 
shown at two radii (corresponding to two green points in optical image): one is at galactic center (top) 
and the other is $r\sim$7 kpc from the center (bottom). 
In each case, we present the spectrum in three wavelength intervals with possible prominent emission lines.
In each panel, the observed spectrum and its error are indicated by the black line and the yellow shaded area, respectively. The best-fitting continuum spectrum is shown in red. 
 }
  \label{fig:spectral fittings}
\end{figure*}

To reduce the statistical dependence between neighboring bins, and to improve the signal-to-noise 
ratio (SNR) of the spectra, we have radially binned our spectra with a bin size of 1.2 arcsec, 
which is roughly equal to the median FWHM of the seeing. 

To measure the emission- and absorption-based parameters, we adopted a chi-squared 
minimization fitting method \citep{Li-05}.
The templates are galactic eigenspectra constructed with a principal component analysis method. 
When fitting the spectrum, we carefully masked the emission lines iteratively. This fitting 
code is efficient and stable, especially for spectra with low SNR. The disadvantage
of our code is that it is unable to fit the line-of-sight (LOS) velocity automatically. 
To overcome this limitation, we used a fixed velocity given by the spectroscopic
 SDSS redshift in the first fitting, 
and extracted the velocities of the emission lines in the red-arm. 
Specifically, we used the H$\alpha$ emission when it was prominent.
Then we fitted the galaxy rotation curve using the formula from \cite{Bohm-04}:
\begin{equation}
V(r) = V_{\rm max}\frac{r}{(r^a+r_0^a)^{1/a}}+V_{0},
\end{equation}
where $r$ is the radius, V$_{\rm max}$ is the maximum rotation velocity, 
V$_{0}$ is a constant offset in velocity,
and $a$ and r$_{0}$ define the shape of the profile. In our procedure, 
we set $a$ to 5 \citep[recommended by][]{Bohm-04} and V$_0$ to 0 km s$^{-1}$. 
Then we obtained the best-fitting result with V$_{\rm max}$=175 km s$^{-1}$ and $r_0$=4.4 arcsec. 
By using this model, we interpolated the rotation velocity at any position along the slit in a 
numerically stable way. 
We then used these interpolated velocities as input and refitted stellar continuum. 
Note that due to the wide slit (wider than the seeing), the precise rotation curve was difficult to 
obtain \citep[see figure 1 in][]{Carton-15}. However, this model provides a reasonable approximation
 to the true velocity curve along the spectroscopic slit. 
      
The right panels of Figure 1 illustrate two examples of spectral fitting result
 at two radii (marked as green points in the optical image): 
one at galactic center (top) and the other $r\sim$7 kpc from the center (bottom). 
In each case, we present the spectrum in three wavelength intervals, where
the observed spectrum and its error are shown as the black line and the yellow shaded area, respectively. 
The best-fitting continuum spectrum is shown in red, where the continuum models match 
very well with the spectral features in all three wavelength intervals.
We compared the central spectrum from the long-slit observation with the SDSS 3-arcsec fiber spectrum, 
and found the two are highly consistent with each other. There are almost no emission line features at the galactic center, 
while significant H$\alpha$ and H$\beta$ emission are seen at the radius of $\sim$7 kpc.  

\subsection{Physical properties of the galaxy}

The physical quantities considered in this work include the stellar
mass $M_{*}$, stellar surface mass density $\mu_{*}$, \sersic\ index $n_{\rm Sersic}$, 
the \nuvr\ color and the global star formation rate (SFR).
The stellar mass, \sersic\ index and \nuvr\ color are obtained from NASA Sloan
Atlas \citep[NSA;][]{Blanton-11}. The stellar surface density is defined as
$\mu_{*}=M_{*}\,(2\pi R_{\rm 50,z}^2)^{-1} $, where
$R_{\rm 50,z}$ is the physical radius that contains half the total light in the $z$-band. 
The global SFR is taken from both MPA-JHU database based on
 SDSS 3-arcsec fiber \citep{Brinchmann-04}, and from \citet{Chang-15} based on multi-band SED
 fittings of photometric data.

The H{\sc{I}} mass is calculated as
$M_{\rm H{\sc I}}$=$2.356\times10^5$($D_{\rm lum}\,{\rm Mpc}^{-1}$)$^2$($F_{\rm tot}\,{(\rm Jy\,km\,s^{-1}})^{-1}$),
where $D_{\rm lum}$
is the luminosity distance
and $F_{\rm tot}$ is the integrated H{\sc{I}} flux.

We used the 4000 \AA\ break (\dindex), H$\delta$ absorption (\ewhda) and H$\alpha$ emission (\ewhae)
 indices to indicate the recent star formation history along the major axis of NGC 6145 
\citep[$<$2 Gyr;][]{Bruzual-Charlot-03, Kauffmann-03, Kriek-11, Li-15}. 
We measured \dindex\ and \hda\  index from the
emission-line-subtracted spectra for each spectrum, while we measured
the emission lines from the stellar-component-subtracted
spectrum by fitting a Gaussian profile to these lines.

We estimated the SFR surface density along the major axis with the following formula 
\citep{Kennicutt-89,Kennicutt-Evans-12}:
\begin{equation}
SFR = \frac{L(H\alpha)}{1.86\times10^{41}\ \rm erg\  s^{-1}},
\end{equation}
where $L(H\alpha)$ is the H$\alpha$ luminosity. 
When calculating the SFR, we corrected the intrinsic dust attenuation using the Balmer decrement \citep{Domnguez-13}. 
We adopted the dust attenuation curve from \cite{Calzetti-00} and assumed an intrinsic flux ratio of $H\alpha/H\beta$=2.86, 
which corresponds to a temperature $T$=10$^4$ K and an electron density $n_e$=10$^2$ cm$^{-3}$
 for Case B recombination. 
We restricted our analysis to spectra for which the SNR of the
 H$\alpha$ and H$\beta$ fluxes was greater than 3.
By using this method, the calculated E(B-V)$_{\rm gas}$ has a large range and varies from 0.0 to 1.3 mag. 
The dust attenuation near the galactic center is small, but becomes
extremely high on the outer regions. 

\section{Results}

\subsection{HI intensity map of NGC 6145}

\begin{figure*}
  \begin{center}
    \epsfig{figure=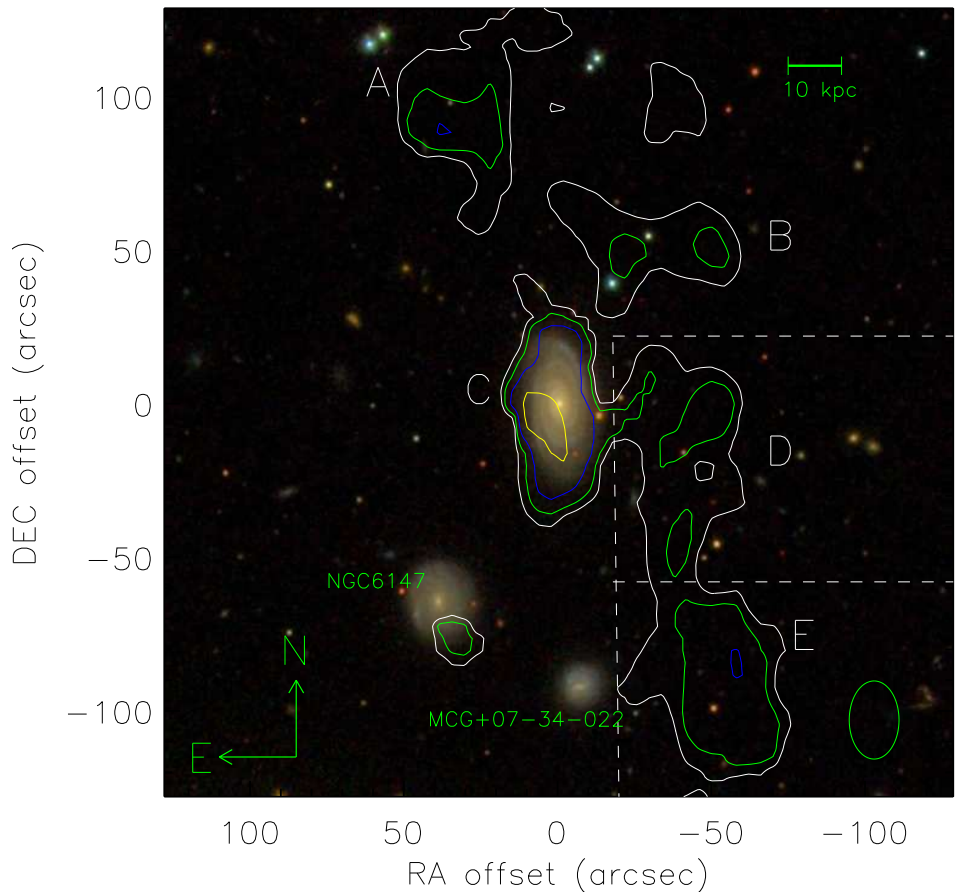,clip=true,width=0.7\textwidth}
  \end{center}
  \caption{ \HI\ column density contours overlaid on the SDSS optical image of NGC 6145. The white, green, blue and yellow 
\HI\ contours represent 0.75, 1.5, 3.0 and 6.0 times 10$^{20}$ atoms cm$^{-2}$, respectively. 
We broadly separated the \HI\ gas into five regions by using the outermost contour, and marked them as A-E, respectively.
We named these five \HI\ regions Clouds A-E in the analysis below.
Since Clouds C, D and E are connected, we separated them with white dashed lines for clarity.
The shape of the beam is plotted in green in the bottom-right corner. 
The map has a size of 147$\times$147 kpc$^2$, and the scale is displayed in the top-right corner.
The map is displayed as north up and east left.}
  \label{fig:HI column density map}
\end{figure*}

Figure 2 shows the contours of \HI\ column density overlaid on the SDSS optical image of NGC 6145. 
The white, green, blue and yellow \HI\ contours represent 0.75, 1.5, 3.0 and 6.0 times 10$^{20}$ atoms cm$^{-2}$, 
respectively. The shape of the beam is plotted in green in the bottom-right corner. 
The map has a size of 147$\times$147 kpc$^2$, and the scale is displayed in the top-right corner.
For analysis convenience, we roughly separated the \HI\ gas into five regions, based on the outermost 
contour, and marked them as A, B, C, D and E.  We named these five \HI\ regions 
Clouds A, B, C, D and E in our analysis below.
Since Clouds C, D and E are connected, we separated them with white dashed lines according to the 
\HI\ distribution. 

As shown in Figure 2, the \HI\ distribution of NGC 6145 is well resolved, 
since the size of \HI\ gas is far greater than beam size.
The \HI\ gas distribution is quite asymmetric: there is almost no \HI\ gas beyond the 
stellar disk to the southeast of the galaxy, while several large \HI\ gas clouds can be seen 
on other sides. These gas clouds, and the diffuse 
 \HI\ gas between them, form a long \HI\ filament with a projected length of $\sim$150 kpc.
On the stellar disk, the \HI\ gas distribution is asymmetric. 
The intensity-weighted center of the HI gas anchored on the stellar disk (Cloud C) 
deviates from its optical center by about 3.3 arcsec,
which indicates a gas accumulation at the southeast of the galactic center.
The main part of \HI\ distribution generally has a similar shape to that of the stellar disk, 
and the \HI\ column density drops significantly just beyond the bright part of the stellar disk.
When going from galactic center outwards, the \HI\ column density becomes lower than 
1.0$\times10^{20}$ atoms cm$^{-2}$, at a radius of 17.7 kpc along the major axis, 
which is slightly larger than R$_{25}$ ($\sim$16.3 kpc), defined as the radius at the surface brightness
 of 25 mag arcsec$^{-2}$ in SDSS $g$-band image. 

The total \HI\ mass is 10$^{9.74}$M$_{\odot}$, when we include \HI\ gas both within and outside the stellar disk.
 We separated \HI\ gas into five regions as shown in Figure 2.  
 The \HI\ mass is 10$^{8.82}$,10$^{8.63}$,10$^{9.22}$,10$^{8.85}$ and 10$^{9.03}$M$_{\odot}$
  for Clouds A-E, respectively.
 Among them, Cloud C corresponds to the stellar disk of NGC 6145 and hosts most of the \HI\ gas (29.8\% of the total). 
 Clouds A, B, D and E host significant \HI\ gas contents, which account for 11.2\%, 7.7\%, 
 12.7\% and 19.4\% of the total \HI\ gas, respectively.  
 In addition to these five dense clouds, there is also some diffuse \HI\ emission that 
  accounts for 18.6\% of the total \HI\ gas. 
Two close companions, NGC 6147 and MCG +07-34-022, can be seen in Figure 2. 
MCG +07-34-022 has no detectable HI emission, which is based on a smooth-and-clip algorithm
 \citep{Serra-Jurek-Floer-12}, while NGC 6147 appears to have a small piece of \HI\ distribution 
 (smaller than the beam size) near its
 optical counterpart.  This \HI\ emission is likely a noise peak, since the corresponding \HI\ mass,
 10$^{7.9} M_{\odot}$, is below the $rms$ noise limit ($\sim$ 10$^{8.1}M_{\odot}$).

The \HI\ distribution of NGC 6145 shows a 150 kpc filament on one side. Interestingly, one-sided \HI\ distributions of galaxies are not very rare. 
There are a few other galaxies that also have one-sided \HI\ filaments, such as 
NGC 4027 \citep{Phookun-92}, Arp 304 \citep{Nordgren-97}, 
and Arp 84 \citep{Kaufman-99}. In these cases, the optical counterparts are usually found 
in the density peaks 
of the one-sided \HI\ distribution, suggesting a merger- or interaction-driven origin. 
For the case presented in this work,  no optical counterparts were found in the \HI\ filament, suggesting that 
pure interaction or merger are not sufficient to explain it. 
In some violent interactions and mergers, one or more \HI\ tidal tails can be seen, 
such as  Arp 299 \citep{Hibbard-Yun-99}, NGC 4038 \citep{Hibbard-01} and the Arp 270 system 
\citep{Clemens-99}. In these cases, both the \HI\ and the optical morphology of galaxies exhibit tidal
 tails or disturbed features.  These galaxies are different to NGC 6145, which, along with its companions, 
does not show disturbed signs, as inferred from the SDSS color image (See Figure 3).  
 
 The one-sided \HI\ filaments are not always traced by close companions, such as  NGC 4654 \citep{Phookun-Mundy-95},
 NGC 4330, NGC 4402, NGC 4522 and NGC 4569  in Virgo cluster \citep{Chung-09, Lee-Chung-15}.
  These galaxies appear to have head-tail shapes, with a compressed edge on one side and a long tail on 
 the other side. Such a morphology has been proposed to indicate ongoing ram-pressure stripping.     
 In contrast, NGC 6145 does not show a head-tail shape, and several gaps can be seen between the \HI\ clouds, which are not 
 seen in typical ram pressure stripped cases. 
There are also some \HI\ filaments that are not traced by optical counterparts and do not show a head-tail shape, 
such as the Leo ring in M96 group \citep{Schneider-89, Watkins-14} and the \HI\ filament in NGC 5291 system \citep{Malphrus-97}. The origins and formation of these structures are unclear,
 and the \HI\ gas in these systems may be primordial \citep{Schneider-89}. 

Generally speaking, in contrast with many galaxies showing one-sided \HI\ distributions, 
the \HI\ filament of NGC 6145 is not traced by close companions, nor does it have a typical ram 
pressure stripped shape. There is also no evidence that the \HI\ filament is due to tidal 
interactions, based on the \HI\ and optical images. 

\subsection{Environment of NGC 6145}

NGC 6145 is located to the west of A2197 \citep{Wrobel-88}, which contains more than 300 galaxy members
 \citep{Yang-07}. The halo mass of this cluster is 10$^{14.6}$ M$_{\odot}$, with a size of $r_{180}\sim$2.8 Mpc, 
where $r_{180}$ is the radius where dark matter halo has an overdensity of 180. 
The Abell 2197 is an X-ray poor cluster \citep{Voges-99}, whose total mass is nearly one third of 
Virgo cluster \citep{Fouque-01}.
The projected distance (D$_{p}$) from NGC 6145 to the cluster center is 1.2 Mpc, and 
it has three close companions within $D_{p}<$150 kpc: NGC 6146, NGC 6147 and MCG +07-34-022.
 
Figure 3 shows the local environment of NGC 6145. Similar to Figure 2,  the \HI\ column density contours 
are displayed in white, green, blue and yellow. The red contours show the radio continuum flux 
at 1.4 GHz observed by WSRT. 
The optical image is centered at NGC 6146, which has a size of $\sim$340 kpc on each side 
at a redshift of 0.029. We labeled the names of the companion galaxies in green.

We listed the basic information of these four galaxies in Table 1. 
From left to right, the columns show the name,  
right ascension, declination, redshift, stellar mass, surface stellar mass density, \nuvr, \sersic\ index $n_{\rm Sersic}$,
SFR from MPA-JHU database, SFR from \citet{Chang-15} and the specific SFR (sSFR)
 \citep{Chang-15}, respectively.
Here we listed SFR obtained by two methods: the first one was derived from the SDSS central 
 3-arcsec fiber spectrum, which reflects the star formation status in the galactic center \citep{Brinchmann-04}; and 
 the second was computed via SED fitting using both SDSS and WISE broad-band fluxes, 
 which represents the star formation status of the whole galaxy. 
 
\begin{deluxetable*}{cccccccccccc}
\renewcommand{\arraystretch}{1.4}
\tablewidth{0pc}
\tablecaption{Physical properties of NGC 6145 and its three close companions. }
\tabletypesize{\footnotesize}
\tabletypesize{\tiny}
\tablehead{    
\colhead{Name} & \colhead{ra} & \colhead{dec} & \colhead{z} & \colhead{$\log M_*$} & \colhead{$\log \mu_*$} & \colhead{NUV$-$r}  & \colhead{$n_{\rm Sersic}$} & \colhead{$\log$SFR$^1$} & \colhead{$\log$SFR$^2$}  & \colhead{$\log$sSFR$^2$} \\
\colhead{ } & \colhead{(J2000)} & \colhead{(J2000)} & \colhead{ } & \colhead{$\rm (M_{\odot})$} & \colhead{($\rm M_{\odot} kpc^{-2}$)} & \colhead{(mag)}  &  \colhead{ }  & \colhead{($\rm M_{\odot} yr^{-1}$)} & \colhead{ ($\rm M_{\odot} yr^{-1}$)}  & \colhead{(\rm yr$^{-1}$)} \\
}
\startdata

NGC 6145        &16$^h$25$^m$02.36$^s$ & +40$^{\circ}$56$^{'}$47.8$^{''}$ &   0.02867 &  10.89   &   8.79  &    4.06  &    1.77  &  -0.76  &  -0.27 &  -11.16\\
NGC 6146        & 16$^h$25$^m$10.33$^s$ & +40$^{\circ}$53$^{'}$34.3$^{''}$&   0.02942 &  11.61  &     9.31   &   5.64  &    5.84  &   -   &  -4.27 & -15.88 \\
NGC 6147        & 16$^h$25$^m$05.84$^s$ & +40$^{\circ}$55$^{'}$43.5$^{''}$ &   0.02894 & 10.30    &  8.21   &   3.23  &   0.96   &  -1.10   &  -2.32 & -12.62 \\
MCG +07-34-022 & 16$^h$25$^m$01.78$^s$ & +40$^{\circ}$55$^{'}$15.6$^{''}$ &   0.02954 & 9.75   &   8.08   &   2.73    &  2.30   & -0.50 &  -0.97 & -10.72\\
\enddata
 \label{tab:sample}
 \tablecomments{ From left to right, columns show the galaxy name, 
  right ascension, declination, redshift, stellar mass, surface stellar mass density, 
NUV$-$r, \sersic\ index $n_{\rm Sersic}$, star formation rate from MPA-JHU database, star 
formation rate from \citet{Chang-15} and the specific star formation rate \citep{Chang-15}, respectively.\\
}
\end{deluxetable*}

\begin{figure*}
  \begin{center}
    \epsfig{figure=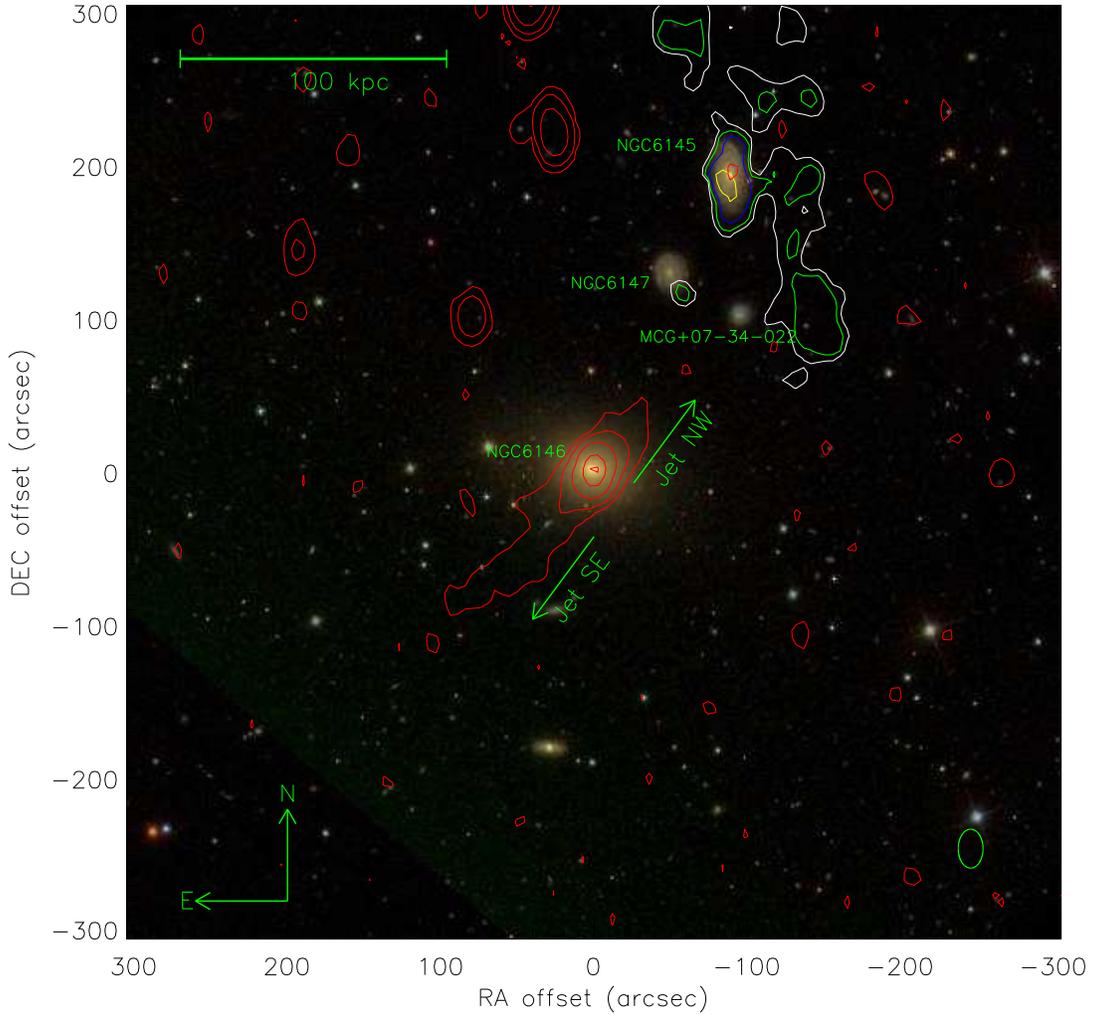,clip=true,width=0.8\textwidth}
  \end{center}
  \caption{Environment of NGC 6145. The optical image shows a large region centered at NGC 6146, which corresponds to
 $\sim$340$\times$340kpc$^2$ at a redshift of 0.029.  Some nearby galaxies can be seen, where  
  their names are labeled with green characters (see Table 1). As in Figure 2,
  the \HI\ column density contours are shown in white, green, blue and yellow.
   The red contours show the radio continuum at 20 cm. 
  The beam shape is plotted at the bottom-right corner in green.  }
  \label{fig:HI and radio emission}
\end{figure*}
 
NGC 6146 is a massive elliptical galaxy with a stellar mass of 10$^{11.6} M_{\odot}$, which is 
about five times massive than NGC 6145. It has no, or very weak, star formation activities at a projected 
distance of 121.8 kpc from NGC 6145. 
NGC 6147 and  MCG +07-34-022 are smaller galaxies, with masses less than 10$^{10.5}M_{\odot}$. 
The line-of-sight velocity difference of NGC 6146, NGC 6147, and MCG +07-34-022 with respect to NGC 6145
is 225, 81 and 261 km s$^{-1}$, respectively, which is much smaller than the velocity 
dispersion of A2197 ($\sim$639 km s$^{-1}$
from \cite{Yang-07} ).  These values indicate that they are physically associated and likely
 gravitationally bound.
There are also two galaxies at a similar redshift in the south of NGC 6146 in Figure 3:
 SDSS J162512.42+405201.6 and SDSS J162512.80+405031.4.
 However, these two galaxies are not included in Table 1, because they are not massive 
  (logM$_{*}$/M$_{\odot}<$10.4) and they are far from the targeted galaxy NGC 6145 with $D_{p}>$ 170kpc.

As shown in Figure 3, NGC 6146 has strong radio emission, with a radio flux density of 155.3 mJy, 
indicating that it hosts a typical ``radio-mode'' AGN.
 When an AGN is at a low level in massive elliptical galaxies, the feedback is usually in a form of radio mode, 
 which uses the mechanical energy of radio-emitting jets \citep{Fabian-12}.
 The radio luminosity of NGC 6146 at 1.4 GHz is 3.08$\times$10$^{23}$ W Hz$^{-1}$, which puts it in the FR I type 
 classification \citep{Ghisellini-Celotti-01}. The jet power $P_{jet}$ we evaluated is $4.3\times10^{43}$ erg s$^{-1}$, 
 by adopting the empirical function from \cite{Cavagnolo-10}.
 
NGC 6146 shows an asymmetric radio-emitting jet along its minor axis,  
with a position angle of 143.5$^\circ$. The radio emission in the northwest has a projected length of $D_p\sim$25 kpc, 
and the other side of radio emission extends to $D_p\sim$60 kpc.
We call the southeast one ``Jet SE'', and the northwest one ``Jet NW'' in the following analysis (see Figure 3). 
 \cite{Wrobel-88} have found that a high-resolution 18 cm image of NGC 6146 shows a $\sim$1 kpc jet knot near 
 the galactic core, and a VLA 20-cm image shows 6-kpc and 11-kpc components at a position angle of 126$^\circ$. 
Compared with the VLA 20-cm image, our 20-cm image has a lower spatial resolution, but higher sensitivity. 
This explains why in our image we are unable to distinguish the 6 kpc component discovered in the VLA image, 
but we can detect faint radio emission out to a distance of at least 60 kpc.

There are several possible reasons that might explain the asymmetric radio jet. One is that the jet is simply intrinsically 
more powerful at one side than at the other side.  
It is also possible that strong ram- or thermal-pressure gradients in its intergalactic medium have twisted, 
or disrupted the radio jet on the weaker side \citep{Henriksen-Vallee-Bridle-81}.  
The asymmetric radio jet is unlikely to be caused by the relativistic beaming effect \citep{Bridle-Perley-84},
because the velocity of the jet  would probably decay to be non-relativistic at a projected distance 
greater than 30 kpc \citep{Strom-83}.


The Jet NW points to the \HI\ gap between Cloud C and Cloud E, as shown in Figure 3. 
This raises an interesting question: is the Jet NW physically related to the observed \HI\ filament? 
 Observationally, \cite{Irwin-87} proposed a possible ram-pressure stripping case of NGC 3073 by 
 the fast, starburst-driven wind from NGC 3079.
 NGC 3073 was found to have a head-tail shape with an elongate \HI\ tail,  which is remarkably 
 aligned with the nucleus of NGC 3079 \citep{Shafi-15}. 
 However, our case is different to the NGC 3073/3079 system. NGC 6145 is larger in optical size 
 and more massive than NGC 3073 (total mass of 1.1$\times10^{9}M_{\odot}$ estimated by \cite{Irwin-87}).
 The feedback from the nucleus of NGC 6146 is much narrower and stronger than the feedback of 
 stellar wind in NGC 3079.  In this sense, one could not rule out the hypothesis
 that the \HI\ filament is due to the Jet NW from NGC 6146, although the \HI\ morphology of NGC 6145 
 appears to be different to that of NGC 3073. 
  
\subsection{Kinematics of \HI\ gas}

\begin{figure*}
  \begin{center}
    \epsfig{figure=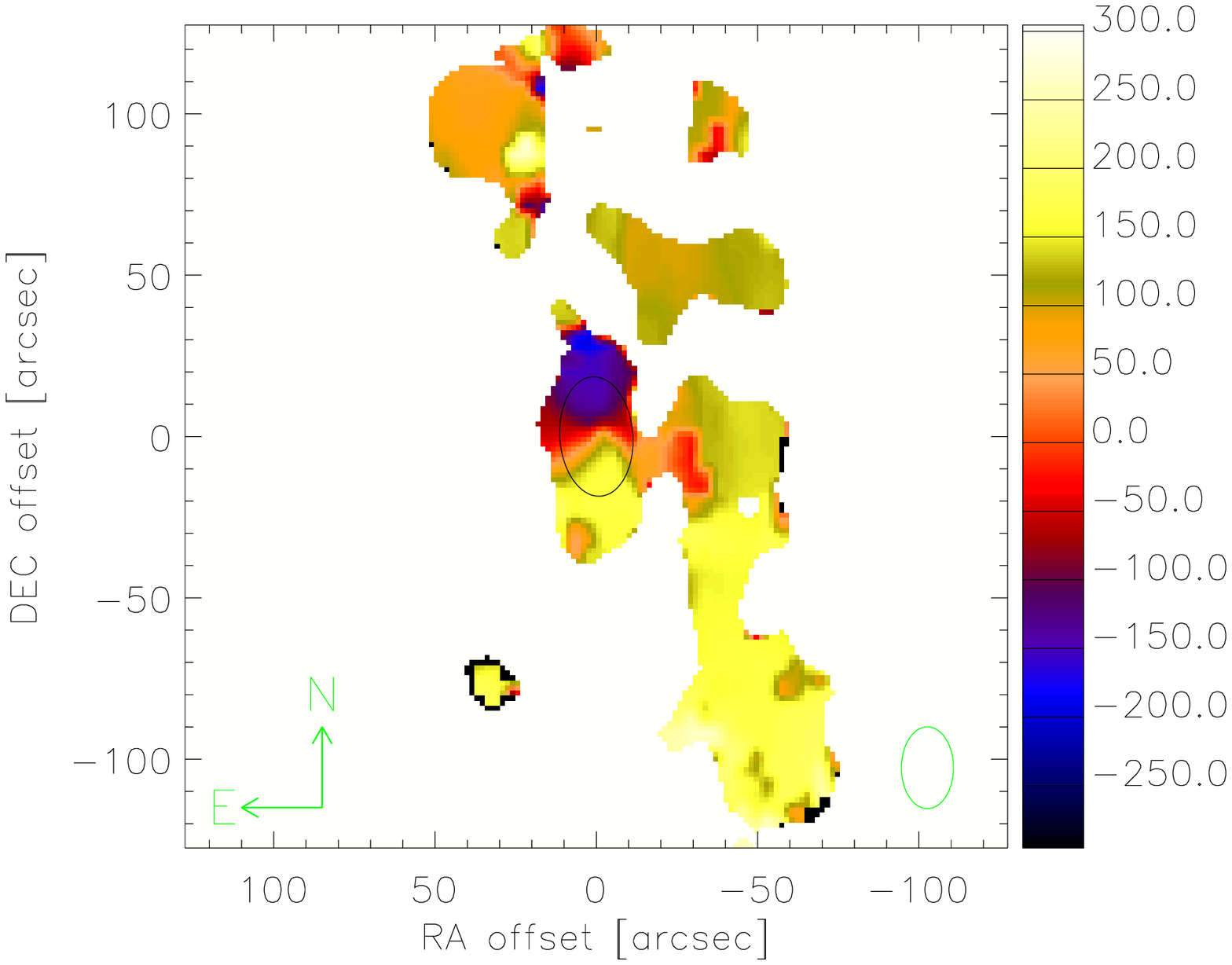,clip=true,width=0.50\textwidth}
    \epsfig{figure=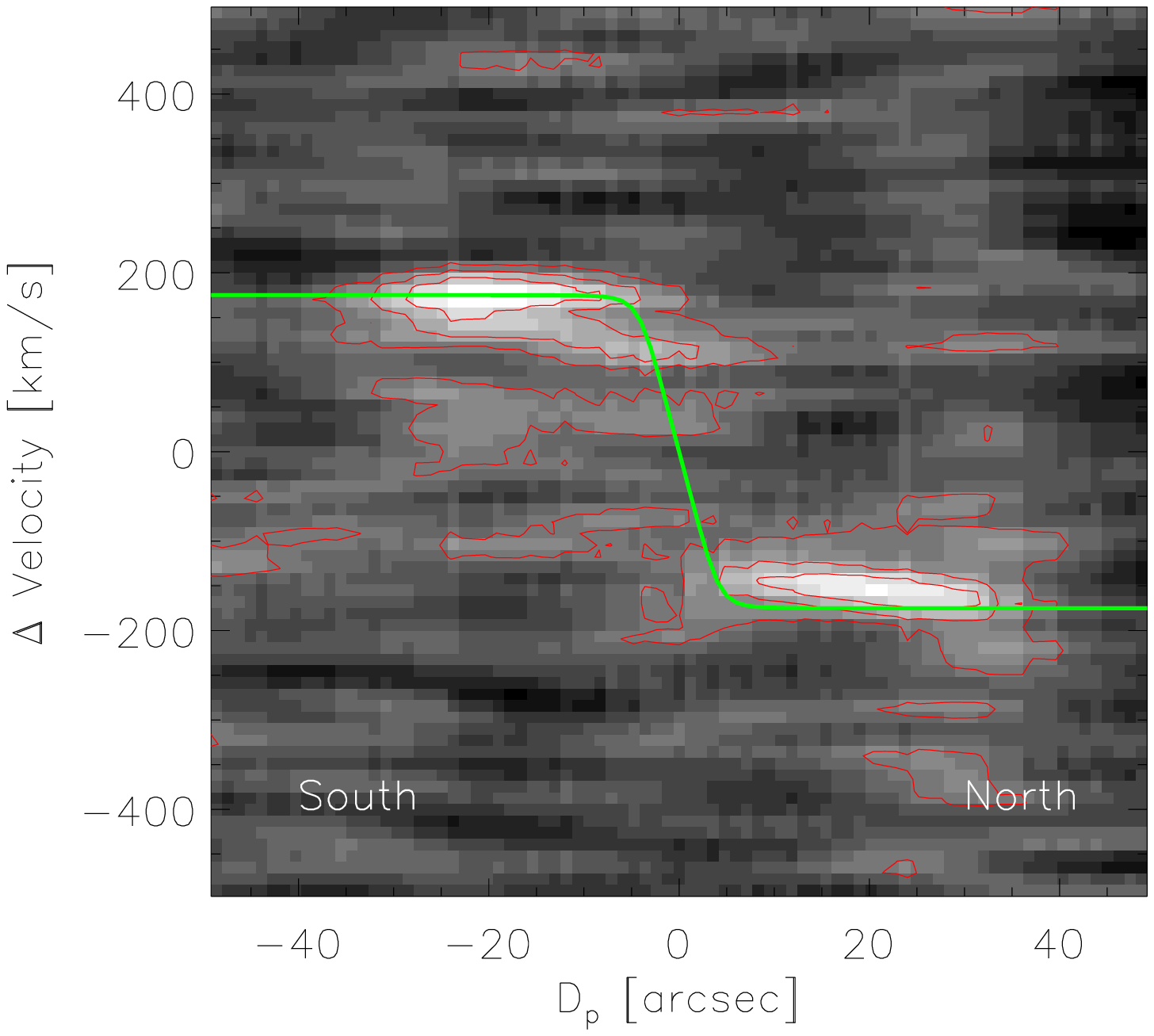,clip=true,width=0.45\textwidth}
  \end{center}
  \caption{Left panel: \HI\ velocity map of NGC 6145. The velocity fields are derived from a Gauss-Hermite fitting
procedure \citep{denHeijer-15}. The central black ellipse shows the stellar disk of this galaxy, where  
the size of ellipse corresponds to a radius that encloses 90\% of the light in $r$-band.
The velocity map is shown by subtracting the central velocity of NGC 6145.
 Right panel: Position-velocity (P-V) diagram of NGC 6145 along the major axis (PA$\sim$4$^\circ$) of 
 this galaxy. The green line is the best-fitting model based on the H$\alpha$ 
 emission from long-slit spectra. The red contours show 2, 4 and 6 times $rms$ noise.}
  \label{fig: HI velocity}
\end{figure*}
  

Supposing that the \HI\ gas far beyond the stellar disk was originally from NGC 6145, and was stripped 
away (ram-pressure stripping or jet stripping),  
one would expect that the velocity of stripped gas clouds should not be 
random. Specifically, these stripped \HI\ clouds should have an overall relative motion along the LOS
 with respect to the host galaxy. 
The left panel of Figure 4 shows the \HI\ velocity map of NGC 6145, which is normalized by subtracting the 
optical-recession velocity.
 The \HI\ velocity fields were derived from a Gauss-Hermite fitting
procedure \citep{denHeijer-15}. The central black ellipse shows the stellar disk of this galaxy, where 
the size of ellipse corresponds to a radius that encloses 90\% of the light in $r$-band (R$_{90,r}$).
The velocity map shows a nice spider diagram in the optical region marked by
black ellipse, which is similar to well-studied cases of non-faceon spirals \citep{Fraternali-02, denHeijer-15}. 
This indicates that the movement of \HI\ gas on stellar disk is not heavily disturbed. 
The mean LOS velocities of Clouds A, B, D and E are 83.4, 115.2, 126.7 and 174.2 km s$^{-1}$,
 respectively, relative to NGC 6145. An increasing LOS velocity is seen along the \HI\ filament 
from Clouds A, B, D to E. The maximum rotation velocity of this galaxy is 175 km s$^{-1}$,
 as determined using Equation 1.  

 The right panel of Figure 4 shows the position-velocity (P-V) diagram of \HI\ gas along the 
 major axis (PA$\sim$4$^\circ$) of NGC 6145.
  The green line shows the bes-fitting rotation curve based on the H$\alpha$ emission 
  from the long-slit spectra (see Equation 1).
 The red contours show the levels of 2, 4 and 6 times $rms$ noise.
 The velocity of the \HI\ gas is consistent with the velocity curve derived from the
 H$\alpha$ emission along the major axis.  

All the gas clouds beyond the stellar disk show positive velocities with respect to the galaxy,
 which indicates that they are moving away from us along the LOS. If the clouds are on the near side of the galaxy, 
 they are moving towards NGC 6145 (accretion) and if they are on the far side of the galaxy, 
 they are moving away from NGC 6145 (stripping).
If we assume accretion is the case, then the \HI\ clouds beyond the stellar disk of NGC 6145 are 
probably coming from NGC 6147 and MCG +07-34-022. Observationally, NGC 6147 and MCG +07-34-022
have no (or false) detection of 21 cm emission (see Figure 3).  
With respect to other interacting systems reported
 in the literature, such as NGC 4027 \citep{Phookun-92}, M81 \citep{Yun-Ho-Lo-94},
 and Arp 84 \citep{Kaufman-99}, 
  the two close companions, NGC 6147 and MCG +07-34-022, do not overlap with the \HI\ filament. 
 This appears to be inconsistent with the explanation that the \HI\ filament is only due to
  the accretion from  NGC 6147 and MCG +07-34-022. Another possible explanation is that the \HI\ filament 
  is the result of accretion from the ram-pressure stripped \HI\ gas originating from the two companion galaxies.
 
Assuming that the \HI\ filament was originally from NGC 6145, the positive velocity of 
gas clouds indicates they are departing from NGC 6145. We speculate that gas Clouds A 
and B originate from the outer regions of stellar disk on the north, and later have been stripped away. 
Cloud D likely originates from the inner region of stellar disk, and Cloud E is likely from the outer region 
of stellar disk to the south, based on positions and velocities of these \HI\ clouds. 
The velocities of the \HI\ in the southern part of the filament are similar
 to those in the southern part of the
 galaxy, while the velocities of the \HI\ in the northern part of the filament are very different
  (by at least 200 km s$^{-1}$) from those in the northern part of the galaxy.
This can be explained by the ram-pressure stripping scenario, where NGC 6145 is moving toward 
us along the LOS with respect to the intergalactic medium, 
although the \HI\ shape is inconsistent with this scenario, as discussed in Section 3.1. 
In this scenario, the gradually changing velocity in the \HI\ filament can be naturally 
explained by the LOS stripping of a rotating \HI\ disk.
In well-studied galaxies with ongoing ram-pressure stripping \citep{Kenney-vanGorkom-Vollmer-04, Chung-09}, 
the kinematic features of \HI\ tails vary from case to case, 
and depend on the strength and direction of stripping,  inclination and \HI\ distribution of stripped galaxy.
In a similar case, NGC 4522 \citep{Kenney-vanGorkom-Vollmer-04} is a nearly edge-on galaxy that is experiencing ongoing ram-pressure stripping. 
The direction of stripping is nearly perpendicular to the stellar disk, as inferred from the projected image. 
The velocity of extra-planar gas in NGC 4522 gradually changes along the direction of the disk, 
which is mainly due to the rotation of the galaxy. 
In our case, the gradually changing velocity in the
\HI\ filament of NGC 6145 can be explained by the combined effect of ram-pressure 
 stripping and rotation, suggesting that the stripping effect is stronger in NGC 6145 than in NGC 4522. 
 
We note that tidal stripping and interaction are also able to cause the gradually changing velocity
 in \HI\ tidal tails and bridges, such as Arp 84 \citep{Kaufman-99} and UGC 12815 in WHISP survey \citep{Swaters-02}, 
 due to the strong tidal forces or gas accretion from close companions, although
 the \HI\ distribution of NGC 6145 does not support these two origins, as discussed in Section 3.1. 
   
\subsection{Star formation histories along the slit}

\begin{figure}
  \begin{center}
    \epsfig{figure=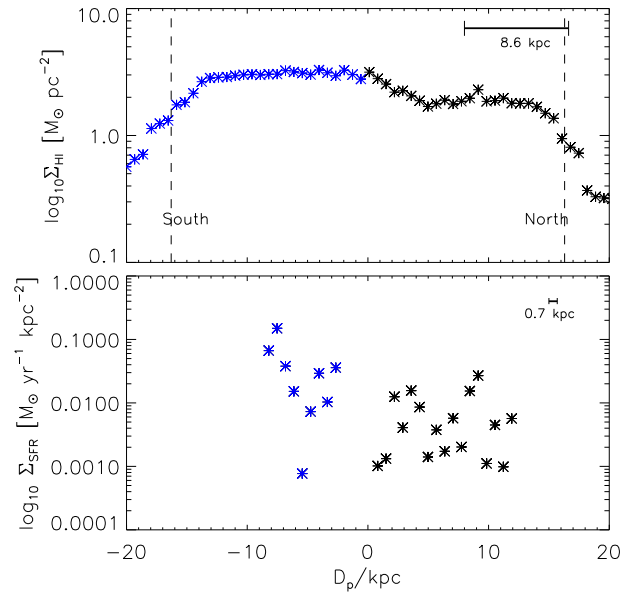,clip=true,width=0.5\textwidth}
  \end{center}
  \caption{Upper panel: Base-10 logarithm of \HI\ gas surface density along the major axis of NGC 6145.
  Bottom panel: Base-10 logarithm of the SFR surface density along the major axis of this galaxy. 
   The blue and black data points represent the two sides around the galactic center.
  In each panel, we have marked out the resolution of \HI\ gas and $\Sigma_{\rm SFR}$ in top-right corner. 
  The $g$-band R$_{25}$ is marked with the vertical dashed lines in the top panel.  
 }
  \label{fig: SFR profile}
\end{figure}

Based on the long-slit spectra, we estimated the SFR surface 
density ($\Sigma_{\rm SFR}$ in units of M$_{\odot}$yr$^{-1}$kpc$^{-2}$) of NGC 6145 
along its major axis with Equation 2 in Section 2.3.  
The bottom panel of Figure 5 shows the base-10 logarithm of $\Sigma_{\rm SFR}$ as a function of 
projected distance to the galactic center.  
H$\alpha$ or H$\beta$ fluxes with SNRs less than 3 were excluded as before.
In the upper panel of Figure 5, we show the \HI\ mass surface density ($\Sigma_{\rm \HI}$ in units of  M$_{\odot}$pc$^{-2}$) 
profile along the major axis. 
In both panels, we labeled the resolution scale of $\Sigma_{\rm HI}$ and 
$\Sigma_{\rm SFR}$ in the top-right corner.

As shown in the bottom panel of Figure 5, $\Sigma_{\rm SFR}$ varies significantly along the 
major axis (by about two orders of magnitude).
The star formation regions are discretely distributed, consistent with the fact that
 the H$\alpha$ emission regions are usually in clumps. 
In the central region ($<2$ kpc), $\Sigma_{\rm SFR}$ is considerably lower than in the outer regions. 
This probably explains why the SFR from MPA-JHU catalog is $\sim$0.5 dex lower than SFR determined by \citet{Chang-15}. 
Note that no data points are seen in the bottom panel beyond $D_p\sim$12 kpc.
 We have visually inspected the spectra beyond $D_p\sim12$ kpc by visually inspection, and 
 did not find any prominent H$\alpha$ emission.

The $\Sigma_{\rm \HI}$ profile shows an almost flat shape ($\sim$2.5 M$_{\odot}$pc$^{-2}$) 
within D$_p<15$ kpc, and starts to drop beyond D$_p\sim$15 kpc (slightly less than the g-band R$_{25}$) on each side. 
 \cite{Wang-16} found a uniform characteristic \HI\ surface mass density ($\Sigma_{\rm \HI, c}$) of
  5.07 M$_{\odot}$pc$^{-2}$ for almost all nearby dwarf and spiral galaxies, 
 measured within the radius where the azimuthally averaged \HI\ reaches 1 M$_{\odot}$pc$^{-2}$.
 The $\Sigma_{\rm \HI}$ of NGC 6145 is lower than $\Sigma_{\rm \HI, c}$ by almost 0.3 dex, which is 
 comparable to the scatter in the \HI\ surface mass densities of nearby dwarf and spiral galaxies,
  thereby raising the possibility that the \HI\ gas in the filament originated from NGC 6145. 
 By comparing  $\Sigma_{\rm SFR}$  (or $\Sigma_{\rm \HI}$) on the two sides, we find, on average,  
 the southern regions 
 have a higher $\Sigma_{\rm SFR}$ and higher $\Sigma_{\rm \HI}$ than the northern regions.

\begin{figure*}
  \begin{center}
    \epsfig{figure=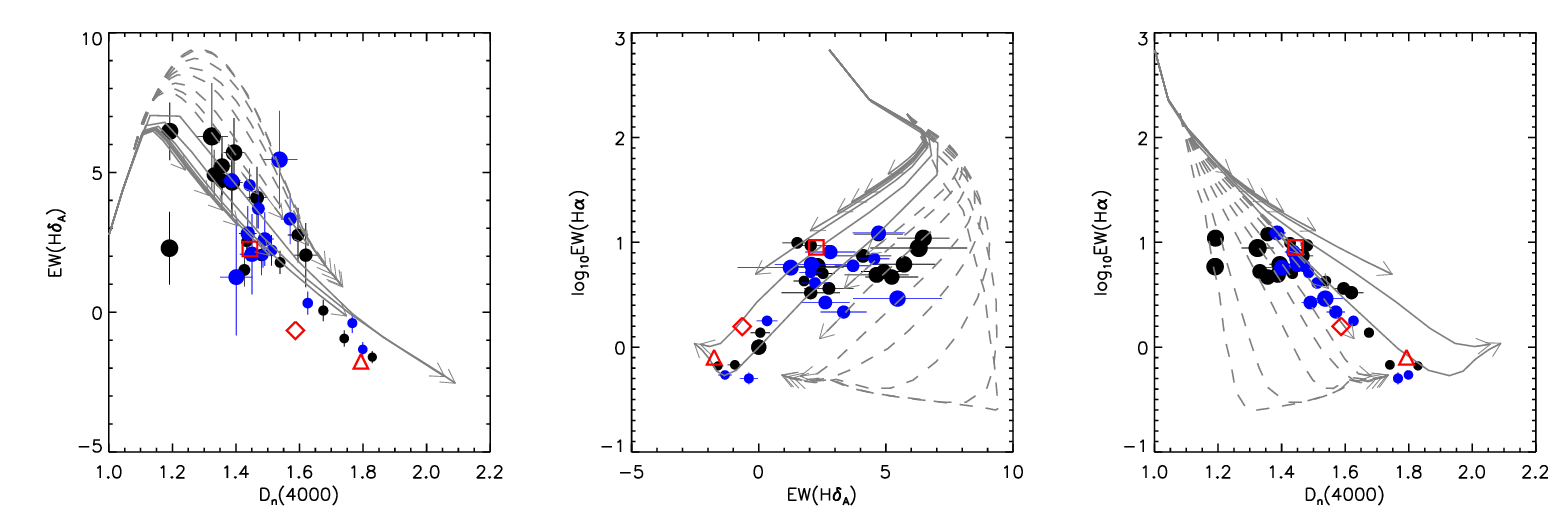,clip=true,width=1.0\textwidth}
  \end{center}
  \caption{Different regions along the major axis of NGC 6145 are shown on the planes of \dindex\ 
vs. \hda\ (left panel), \hda\ vs. \lgewhae\ (middle panel) and \dindex\ vs. \lgewhae\ (right panel).
 Symbol sizes represent distances to galactic center, where larger distances are indicated by
  larger symbol sizes.
 We indicate the southern and northern sides from galactic center with blue and black data points, 
 respectively (see Figure 1), and regions with H$\alpha$ flux SNRs less than 3 are excluded.  
 The solid and dashed lines are solar metallicity models that follow exponentially declining star formation histories 
  (${\rm SFR}\propto\exp(-t/\tau)$) with different $e$-folding times based on the stellar population synthesis 
  code of \cite{Bruzual-Charlot-03}. 
The gray solid lines represent the continuous star formation models with long $e$-folding times
($\tau>5\times10^8$yr), and the gray dashed lines represent the 
star formation bursts with reasonably short $e$-folding times ($\tau<5\times10^8$yr).
Different lines represent the different $e$-folding times. 
 We also plot the MPA-JHU datasets of NGC 6145, NGC 6147 and MCG +07-34-022 on these planes, shown
 as red triangles, diamonds and squares, respectively.
}
  \label{fig: SFH diagram}
\end{figure*}  
  
We used the 4000 \AA\ break (D$_n$(4000)), \hda\ absorption index (\ewhda) and H$\alpha$ emission line 
equivalent width (\ewhae) to trace the recent star formation history \citep{Bruzual-Charlot-03,
Kauffmann-03, Li-15}.
D$_n$(4000) has been commonly used as an indicator of the mean stellar age
for stellar populations younger than 1-2 Gyr, \hda\ index
indicates the star formation that occurred 0.1-1.0 Gyr ago, and \ewhae\ traces
 the strength of very recent (within 50 Myr) or ongoing star formation.

Different regions along the major axis of NGC 6145 are plotted on the planes of \dindex\ 
vs. \ewhda\ (left panel), \ewhda\ vs. \lgewhae\ (middle panel) and \dindex\ vs. \lgewhae\ (right panel), 
as shown in Figure 6.
 Symbol sizes represent projected distances to the galactic center, where larger symbols indicate
  larger distances.
Similar to Figure 6, the blue data points represent the southern regions of NGC 6145 along its major axis,
  while black data points represent the northern regions of NGC 6145 (see Figure 1). 
 We also plotted the MPA-JHU measurements of NGC 6145 on these planes (red triangles).

For comparison, we present \dindex, \hda\ index and \lgewhae\ values by applying the \cite{Bruzual-Charlot-03} (BC03) 
models of solar metallicity with exponentially declining star formation histories (${\rm SFR}\propto\exp(-t/\tau)$).
The gray solid lines represent the continuous star formation models with
$\tau>5\times10^8$yr, and the gray dashed lines represent the 
star formation bursts with $\tau<5\times10^8$yr.
Different lines represent the different characteristic timescales $\tau$. 
The H$\alpha$ luminosity was computed by converting Lyman continuum photons to H$\alpha$ flux following
\citet[][see equations B2-B4 in their appendix]{Hunter-Elmegreen-04}, and taking the
recombination coefficients and flux ratio of H$\alpha$/H$\beta$ from \citet{Hummer-Storey-87}.

As shown in Figure 6, \dindex\, \ewhda\ and \ewhae\ derived from the SDSS 3-arcsec fiber spectrum are 
 consistent with the measurements based on the long-slit spectra at galactic center. 
The central region of NGC 6145 shows very weak or no star formation activities, whose
\dindex\ is about 1.8 and \ewhae\  is less than 1 \AA.  
In contrast to the central region, the outer regions are actively forming stars, since many outer regions
have smaller 4000 \AA\ break values (\dindex$<$1.4) and stronger H$\alpha$ emission (\ewhae$>$2 \AA) .

In addition, some data points are located in the starburst region according to the BC03 models, 
as shown in the three diagrams in Figure 6. 
In a parallel paper, we found that almost all regions of galaxies 
strictly follow the continuous star formation models (BC03) based on the MaNGA \citep{Bundy-15} data. 
This result is attained regardless of stellar mass and quenching status of the galaxy, indicating that 
starburst activities are rare in galaxies with regular morphologies (Wang et al. 2017, in preparation).
However, some outer regions of NGC 6145 are obviously star-bursting, as indicated by
 all three diagrams. 

NGC 6147 and MCG +07-34-022 are also plotted on the diagrams as red diamonds and squares.
As noted above, MCG +07-34-022 is a star-forming 
galaxy with no \HI\ emission, suggesting that it experienced recent gas removal. 
NGC 6147 has very weak star formation activities ($\log$sSFR$\sim-$12.6), and appears to have
 a small piece of \HI\ distribution (smaller than the beam size) near its optical counterpart, which 
 is likely a noise peak. 
We infer that the cold gas of NGC 6147 has probably been removed by external processes \citep{Peng-10}, such as ram-pressure 
stripping. Subsequently, the star formation activity could not be sustained without enough cold gas.

We note that the H$\alpha$ emission is due to H{\sc{II}} regions throughout NGC 6145, at least along its major axis.
In particular, there is no evidence of shock-excited gas or AGN-excited gas, which can be traced  
by strong [OIII]$\lambda$5007 emission lines \citep{Dopita-77}. 
In addition, the flux ratios of [NII]/H$\beta$, [OII]/H$\beta$ or [OI]/H$\alpha$ are too weak to classify it 
as the shock-excited type \citep{Dopita-84, Dopita-Sutherland-95}.
Our long-slit spectra show no or very weak [OIII]$\lambda$5007 emission along the whole slit 
(see two spectral examples in Figure 1). 

\section{Summary and Discussion}

By using the \HI\ datacube from the ``Bluedisk'' project \citep{Wang-13}, 
 we presented and studied the peculiar \HI\ morphology of the galaxy NGC 6145,
 which has an \HI\ morphology that is not obviously induced by tidal or ram-pressure stripping.
The optical long-slit spectra of NGC 6145 orientated along its major axis obtained with 
WHT \citep{Carton-15} were also investigated.

The \HI\ morphology of NGC 6145 is rather asymmetric, with a 150 kpc \HI\ filament located to the west. This filament 
 consists of several \HI\ clouds, as shown in Figure 2. We have compared the \HI\ morphology with other one-sided \HI\ 
distributions presented in the literature, and found the \HI\ distribution is different to the 
typical tidal or ram-pressure stripped HI shape, and it cannot be attributed to a pure accretion event.   
According to the \HI\ distribution, we broadly separated the \HI\ gas into five regions as shown in Figure 2. 
We found that the \HI\ gas on the stellar disk (Cloud C) accounts for 29.8\% of the total \HI\ gas, while the 
Clouds A, B, D and E account for 11.2\%, 7.7\%, 12.7\% and 19.4\%, respectively. 
 Furthermore, we analyzed the kinematics of the \HI\ gas and found that Clouds A, B, D and E all have 
 positive velocities with respect to NGC 6145.
 The velocity of the \HI\ gas beyond the stellar disk gradually increases along the \HI\ filament from Clouds A to E. 
 To investigate the underlying physical mechanisms that account for the \HI\ filament, 
 we investigated the local environment of NGC 6145 and found three close companions at a similar 
 redshift: NGC 6146, NGC 6147 and MCG +07-34-022. 
 NGC 6147 and MCG +07-34-022 are much less massive than NGC 6145, while NGC 6146 is a 
 more massive elliptical galaxy with extended radio emission. The direction of the radio jet roughly
  points to the \HI\ gap between Clouds C and E in NGC 6145.  
 
 We also used the long-slit spectra of NGC 6145 along its major axis to study its star formation history.
 We radially binned our spectra with a bin size of 1.2 arcsec, 
 and performed spectral fitting to derive the emission- and 
 absorption-based parameters. We used the 4000\AA\ break, H$\delta$ absorption index and H$\alpha$ emission
 line equivalent width to trace the recent star formation histories \citep{Bruzual-Charlot-03, Kauffmann-03}. 
By investigating the diagrams of \dindex\ vs. \ewhda, \ewhda\ vs. \lgewhae\ and \dindex\ vs. \lgewhae,
we found some outer regions along the major axis reside in the starburst region on the quenching diagnostic diagrams,
while almost no star formation activities were found in galactic center ($<$2 kpc).

There are some possible mechanisms that can account for the \HI\ filament in NGC 6145, including 
accretion from nearby companions (NGC 6147 and MCG +07-34-022), 
 tidal stripping by NGC 6146, ram-pressure stripping, the stripping by Jet NW and the combined effects of 
 two or more processes.  

The one-sided \HI\ distribution in mergers or interactions usually hosts one or more companions, 
who trace the HI column density peaks \citep[e.g.][]{Phookun-92, Nordgren-97}. 
However, in our case, two close companions of NGC 6145 
do not reside in the \HI\ filament, and they do not show disturbed features in SDSS image. 
This argues against the scenario 
that the \HI\ filament formed only by the accretion from NGC 6147 and MCG +07-34-022. 
The shape of \HI\ filament appears to be different to the typical tidal stripped \HI\ tails, 
which are usually distributed roughly symmetrically on both sides of galaxy 
\citep{Odenkirchen-01, Klimentowski-07, Rodruck-16}, 
along with disturbed optical morphology and enhanced star formation in galactic center \citep{Fujita-98, Li-08, Smith-Davies-Nelson-10}.
NGC 6145 has enhanced star formation activities in the outer region, and not in the galactic center, which is 
inconsistent with the scenario that the \HI\ filament is due to the tidal stripping by NGC 6146. 

The positive LOS velocity of the filament with respect to NGC 6145 and the gradual change 
in the LOS velocity along the filament, can be naturally explained by the combined effects of ram-pressure 
stripping and rotation.
 Ram-pressure stripping is also able to produce the enhanced star formation activities seen in the stellar disk.  
However, the global \HI\ distribution appears to be inconsistent with the ram-pressure stripping picture. 
Assuming that the \HI\ filament is a ram-pressure stripped ``tail'', the opening angle of the \HI\ 
tail is nearly 180 degrees,
 measured with the vertex at the outer edge of the \HI\ disk in the leading side. 
However, the opening angle of \HI\ tail for typical ram-pressure stripped case is much narrower than 
seen here \citep[e.g.][]{Gavazzi-95, Kenney-vanGorkom-Vollmer-04, Chung-09}. 
It is also difficult to explain why Clouds A and E are more than 30 kpc away from the edge of stellar disk, 
along the direction perpendicular to the stripping direction. 
  
 An alternative explanation is that the formation of \HI\ filament is due to the stripping by Jet NW 
 from NGC 6146. It appears that the jet-stripping scenario works better in explaining the 
 global \HI\ shape than ram-pressure stripping scenario,  although the direct 
  evidence for jet-cold gas interaction has not yet been found in our data. 
 One may doubt whether a narrow jet could produce the 150-kpc-scale \HI\ distribution, 
 and particularly Clouds A and B which do not lie along the jet axis. The \HI\ filament may be
 produced under the jet stripping hypothesis if NGC 6145 has been going across the Jet NW 
 from south to north, although this has not been demonstrated by hydrodynamic simulations.
 We note that there are some other \HI\ gaps along the filament, and it is not known whether the
  jet-stripping picture could explain these gaps or not, as there is a lack of examples in the literature
   for a gas-rich disk galaxy being shot by an external narrow jet. 
  
We also roughly estimated the energy 
required to produce the observed distortion in the \HI\ distribution under the jet stripping scenario.
 Here for simplicity we just estimated the changes in the kinetic 
and gravitational energies of the stripped \HI\ clouds under the following assumptions:
1) the velocity of the stripped gas perpendicular to the LOS is similar to its velocity along the LOS
 ($\sim$180 km s$^{-1}$ with respect to NGC 6145);  
 2) the stripped gas clouds originated from the nearest edges of the stellar disk of
 NGC 6145; and 3) the gravitational force on the stripped \HI\ clouds is dominated by NGC 6145 
 instead of the other three companion galaxies.
  Thus the estimated total energy required is $\sim$6$\times$10$^{57}$ erg, which corresponds to 
  a required power of $P_{\rm gas}\sim$10$^{42}$ erg s$^{-1}$ by assuming an interaction timescale of 0.2 Gyr. 
  The interaction timescale of $\sim$0.2 Gyr is estimated from the projected distance between the stripped \HI\ clouds 
  and the stellar disk, $D_p\sim$40 kpc, and the velocity perpendicular to the LOS of 180 km s$^{-1}$. 
  This is consistent with the estimated jet power ($P_{\rm jet}>2P_{\rm gas}$), 
  which indicates that the jet is strong enough to cause such an \HI\ distribution. 
 
 In principle, a combination of early-time ram-pressure stripping of the \HI\ gas from 
 NGC 6147 and MCG +07-34-022, and subsequent accretion onto NGC 6145 could also explain the 
observed \HI\ distribution, while this hypothesis could not be proved by our current observations. 

The system we investigated is a good candidate for radio-jet and cold gas interaction. 
 If the jet stripping scenario is true, this system provides an ideal laboratory to study 
the details of radio jet interactions with a gas-rich disk galaxy. 
However, more observations are needed to prove or disprove the jet-stripping scenario. 


\acknowledgments

We thank David Carton for providing the long-slit spectra. We thank the anonymous referee for useful suggestions. 
This work is supported by the Strategic Priority Research Program 
``The Emergence of Cosmological Structures'' of the Chinese Academy of Sciences (No. XDB09000000), 
the National Basic Research Program of China (973 Program)(2015CB857004), and the National 
Natural Science Foundation of China (NSFC, Nos. 11225315, 1320101002, 11433005,
 11421303, and 11643001). 
EW was supported by the Youth Innovation Fund of University of Science and Technology of China (No. WK2030220019). 
 FG gratefully acknowledges support from Chinese Academy of Sciences through the
Hundred Talents Program. CL acknowledges the support of National Key Basic Research Program of 
China (No. 2015CB857004), NSFC (Grant No. 11173045, 11233005, 11325314, 11320101002) and the Strategic 
Priority Research Program ``The Emergence of Cosmological Structures'' of CAS (Grant No. XDB09000000).
GBM was supported by the Fundamental Research Funds for the Central Universities and the 
National Natural Science Foundation of China (NSFC--11421303).

\bibliography{rewritebib.bib}

\label{lastpage}
\end{document}